\documentclass[12pt]{article}
\newcommand{\be}{\begin{equation}}
\newcommand{\ee}{\end{equation}}
\newcommand{\ba}{\begin{eqnarray}}
\newcommand{\ea}{\end{eqnarray}}

\begin{document}

\title{{\bf Anthropic Estimates for Many Parameters of Physics and Astronomy}
\thanks{Alberta-Thy-1-17, arXiv:1703.03462 [hep-th]}}

\author{
Don N. Page
\thanks{Internet address:
profdonpage@gmail.com}
\\
Department of Physics\\
4-183 CCIS\\
University of Alberta\\
Edmonton, Alberta T6G 2E1\\
Canada
}

\date{2017 April 26}

\maketitle
\large
\begin{abstract}
\baselineskip 18 pt

Anthropic arguments of Carter, Carr, and Rees give two approximate power-law relations between the elementary charge $e$, the mass of the proton $m_p$, and the mass of the electron $m_e$ in Planck units:  $m_p \sim e^{18}$, $m_e \sim e^{21}$.  A renormalization group argument of mine gives $e^{-2} \approx -(10/\pi)\ln{m_p}$.   Combining this with the Carter-Carr-Rees relations gives $e^2\ln{e} \approx -\pi/180$.  Taking the exact solutions of these approximate equations gives values for $e$, $m_p$, and $m_e$ whose logarithms have relative errors only 0.7\%, 1.3\%, and 1.0\% respectively, without using as input the observed values of any parameters with potentially continuous ranges.  One can then get anthropic estimates for the masses, sizes, luminosities, times, velocities, etc.\ for many other structures in physics and astronomy, from atoms to giraffes to the universe, as simple powers of the anthropic estimate for the elementary charge.  For example, one gets an anthropic estimate for the real part of the logarithm of the dark energy density with a relative error of only 0.2\%.

\end{abstract}

\normalsize

\baselineskip 20 pt

\newpage

\section{Introduction}

A goal of physics is to predict as much as possible about the universe.  (Here I mean `predict' in the sense of deducing from theories and assumptions about the universe, whether or not the result of the prediction has been known by observation temporally before the prediction is made.)  One part of this goal would be to predict the observed constants of physics, such as the mass and charge of the proton and of the electron, and the cosmological constant.  A second part would be to make approximate predictions of astronomical parameters, such as the masses, sizes, luminosities, and lifetimes of stars, and the masses, sizes, and temperatures of habitable planets.  A third part would be to make rough predictions of parameters of observers, such as typical sizes for ones in some ways like humans.

The fondest hopes of many physicists would be to find a theory that predicts all the constants of physics precisely, and perhaps also the cosmological parameters.  One might expect that observers would exist for some range of times within the universe and so not expect absolutely precise predictions for $t_{\mathrm{obs}}$.

For a time it was hoped that superstring/M theory would be a predictive theory of this type, ultimately leading to precise predictions of all the constants of physics (since superstring/M theory has no fundamental adjustable dimensionless constants for the dynamical theory, in distinction to such things as vacuum expectation values whose freedom can be considered to be part of the initial or boundary conditions).  Some physicists, such as David Gross, continue to hold out this hope.  However, it has been discovered that superstring/M theory appears to have an enormous landscape of possible vacua \cite{BP,BP2,KKLT,Susskind}, each with different effective constants of physics (what I have simply called `constants of physics' above and shall continue to do, since if superstring/M theory is correct, there are no fundamental true constants of physics other than what can in principle be deduced from mathematical constants).  Therefore, superstring/M theory by itself may not give unique predictions for the constants of physics.

If the constants of physics turn out to be analogous to cosmological parameters in that they are determined by initial conditions, it might seem rather hopeless to try to predict them, unless one can get a definite theory for the initial conditions.  However, the superstring/M landscape appears to have the property that there can be transitions between huge sets of the different vacua, so that perhaps some simple sets of initial states can lead to fairly definite distributions of vacua and hence of the sets of constants of physics.  This could then lead to predictions of the statistical distribution of the sets of constants of physics.  Nevertheless, this distribution is complicated by the fact that different vacua are expected to lead to different numbers or different distributions of observers and observations, so that the statistical distribution of observations has an observership (or `anthropic') selection effect that modifies the original distribution of the sets of constants of physics.  There is the further complication that the numbers of observations for each vacuum can be infinite, leading to the necessity of performing some regularization of the results and the corresponding `measure problem' \cite{measure}.  There are many competing proposals for solving the measure problem which lead to different statistical distributions of the sets of constants of physics, but so far no single proposal is so compelling that it has become universally accepted.

Here I do not wish to go deeper into this controversial issue but point out how one can use anthropic arguments for approximate relations between the mass and charge of the proton and of the electron to get definite approximate predictions for their values, and then further anthropic arguments can be used to get approximate values for many other parameters of physics, astronomy, and even biology.  These approximate values can be obtained from purely mathematical equations, using no input from observed parameters that are other than integers (such as the number of generations of quarks and leptons, and the number of dimensions of space, which are not yet predicted by these arguments).  Because constants like the cosmological constant are more than a hundred orders of magnitude away from Planck values, it is far too much to expect the approximate relations to give predictions with small relative errors for the quantities themselves, but for the logarithms the predictions are very close to the observed values, generally within the order of a percent.

\section{Planck units}

To avoid the historical accidents of most conventional human units, here I shall express physical parameters as dimensionless multiples of Planck units defined in terms of the speed of light $c$, Planck's reduced constant $\hbar$, Newton's gravitational constant $G$, Coulomb's electric force constant $(4\pi\epsilon_0)^{-1}$, and Boltzmann's constant $k_B$ as \cite{NIST}

Planck mass $\sqrt{\hbar c/G} = 2.176\,470(51)\times 10^{-8}$ kg,

Planck length $\sqrt{\hbar G/c^3} = 1.616\,229(38)\times 10^{-35}$ m,

Planck time $\sqrt{\hbar G/c^5} = 5.391\,16(13)\times 10^{-44}$ s,

Planck charge $\sqrt{4\pi\epsilon_0\hbar c} = 1.875\,545\,956(41)\times 10^{-18}$ C,

Planck temperature $\sqrt{\hbar c^5/G}/k_B = 1.416\,808(33)\times 10^{32}$ K,

Planck energy $\sqrt{\hbar c^5/G} = 1.956\,114(45)\times 10^9$ J $= 1.220\,910(29)\times 10^{19}$ GeV,

Planck power $c^5/G = 3.628\,37(17)\times 10^{52}$ W,

Planck energy density $c^7/(\hbar G^2) = 4.633\,25(44)\times 10^{113}$ J/m$^3$\\ $= 2.891\,85(27)\times 10^{132}$ eV/m$^3 = 0.849\,625(80)\times 10^{200}$ eV/Mpc$^3$. 

The digits in parentheses at the end of the numerical values that precede the powers of 10 are the uncertainties in the corresponding number of preceding digits.  Planck units other than the Planck charge involve powers of Newton's gravitational constant $G$, which has a relative standard uncertainty of $4.7\times 10^{-5}$ that dominates the uncertainties of these Planck units \cite{NIST}.

\section{Observed Physical and Astronomical Parameters}

We can now express some of the observed constants of physics \cite{NIST} that are most important in cosmology and astronomy as follows in conventional units, followed by their values in Planck units:

Elementary charge $e = 1.602\,176\,6208(98)\times 10^{-19}$ C\\ 
$= \sqrt{\alpha} = [137.035\,999\,139(31)]^{-1/2} = [11.7062376167(13)]^{-1} = 0.085\,424\,543\,1148(98)$.  

A possible mnemonic is that $(1 + 1/160\,000\,000)/137.036$ is within the present experimental uncertainty for the fine structure constant $\alpha = e^2$, which has a relative uncertainty of $2.3\times 10^{-10}$, 200\,000 times smaller than the relative uncertainty in Newton's gravitational constant $G$.

Note that here and henceforth I shall always use $e$ as the value of the elementary charge and {\it not} for the base of the natural logarithms.  If I wish to write the exponential of $x$, in this paper I shall write it as $\exp{(x)}$, and not as $e^x$, since here $e^x$ will always mean the value of the elementary charge raised to the power $x$.

Proton mass $m_p = 1.672\,621\,898(21)\times 10^{-27}$ kg $=0.938\,272\,0813(58)$ GeV/c$^2$\\$ = 7.68502(18)\times 10^{-20} = e^{17.8903452(96)} = [1.309656(31)]e^{18} = [0.999987(23)] 2^{-26} 3^{2} 5^{10} e^{18}$. 

Another expression is $m_p = 1.002\,420(23)\times 2^{-127/2}$, for which an approximation within the present uncertainty is $1.0024\times 2^{-127/2}$; the second factor is experimentally indistinguishable from the inverse square root of $2^{127}-1$, the largest number ever shown by hand to be prime, by \'{E}douard Lucas in 1876, a record that is likely to last forever among humans on Earth.  The factor of 1.0024 may be replaced (and slightly improved) by the small musical interval known as the vulture comma \cite{vulture}, $2^{24} 3^{-21} 5^{4} = (1/3)(320/243)^4 \approx 1.002\,428\,866$.  Therefore, within the experimental uncertainties, the proton mass in Planck units is the vulture comma divided by the square root of the largest number ever shown by hand to be prime.

Electron mass $m_e = 9.109\,383\,56(11)\times 10^{-31}$ kg $=0.000\,510\,998\,9461(31)$ GeV/c$^2$\\
$ = 4.18539(10)\times 10^{-23} = e^{20.9452458(96)} = [1.144196(27)]e^{21} =  [0.999982(23)] 2^{-67} 3^{38} 5^3 e^{21}$.

Mnemonic values for the electron mass that are within one standard deviation of the experimental uncertainties are $\sqrt{m_p}\alpha^6/1.00018$ and also $\sqrt{m_p}\alpha^6/1.000188$, where $1.000188 = 2^{-4} 3^6 5^{-6} 7^3 = (1/2) (1.26)^3$ is known as the landscape comma \cite{landscape}, with 1.26 being what one iteration of Newton's method gives for the cube root of 2 when one starts with the approximation $5/4 = 1.25$, which itself is the basis for the fact that a kilobyte has roughly a thousand bytes and the fact that a musical interval of four semitones in equal temperament is fairly close to a perfect major third with simple rational ratio 5/4.  Therefore, within the experimental uncertainties, the electron mass in Planck units is the square root of the proton mass multiplied by the sixth power of the fine structure constant and divided by the landscape comma.  Other than the correction by the division by the landscape comma, this is the result of the rather remarkable coincidence (but surely just a coincidence) that $m_e = (\sqrt{m_p}\alpha^6)^{1.00000350(46)}$, with the exponent very near unity, but still almost 8 standard deviations away from unity in terms of the present observational data.  

To get a relationship that is within the uncertainties without using the landscape comma, one may introduce the hydrogen atom mass $m_H$ and replace the landscape comma by the cube root of the hydrogen-to-proton mass ratio, $(m_H/m_p)^{1/3} \approx 1.0001815$, so that (here restoring the constants $G$, $\hbar$, and $c$)

$\left(\frac{1}{\alpha}\right)^{72} \left(\frac{Gm_e^2}{\hbar c}\right)^3
\left(\frac{m_e}{m_p}\right)^6 \left(\frac{m_H}{m_p}\right)^4 = 1.000016(140)$.

Bohr radius $a_0 = \hbar/(m_e c\alpha) = m_e^{-1} e^{-2} = 0.529\,177\,210\,67(12)\times 10^{-10}$ m\\ $= 3.274\,147(75)\times 10^{22}$.

Within its experimental uncertainties, the Bohr radius in Planck units is the landscape comma divided by the square root of the proton mass (which can be taken to be the vulture comma divided by the square root of the 12th Mersenne prime) and divided by the seventh power of the fine structure constant (which for this purpose can be taken to be 1/137.036).

Density of water $\rho_{H_2O} = 999.9720$ kg/m$^3$\\
$ = 1.939\,739(91)\times 10^{-94}$.

Age of universe $t_0 = 13.80(4)$ Gyr $= 5.040(14)$ trillion days $=4.355(13)\times 10^{17}$ s\\ $= 8.08(2)\times 10^{60}$.

A mnemonic approximation well within the current uncertainty for the age of the universe in Planck units is $e^{-57}/0.98304$, where $e = \sqrt{\alpha}$ is, as always in this paper, the elementary charge in Planck units, and $1/0.98304 = 2^{-10} 3^{-1} 5^5 = 3125/3072 \approx 1.0172526$ is the magic comma \cite{magic}.  Therefore, within the observational uncertainties, the age of the universe is the magic comma divided by the 57th power of the elementary charge, which gives 13.79733(32) Gyr, where here the uncertainty given for this number of gigayears comes mainly from the $4.7\times 10^{-5}$ relative uncertainty of $G$ in the Planck time; the uncertainty in the elementary charge $e$, which is less than one part in eight billion, is negligible in comparison.  For this calculation, replacing $e$ by $(137.036)^{-1/2}$ gives a negligible change of the approximation for the age of the universe, then becoming 13.79734(32) Gyr.

Cosmological constant $\Lambda = [0.998(32)]$ (10 Gyr)$^{-2}$
= [10.01(16) Gyr]$^{-2} = 1.002(32)\times 10^{-35}$ s$^{-2}$\\
$ = 2.91(9)\times 10^{-122} = [0.998(32)](3\pi\,5^{-3}2^{-400}) = [1.837(59)] e^{114} \approx$ ten square attohertz.

Therefore, within its observational uncertainty, the cosmological constant is $3\pi$ divided by $5^3 2^{400}$, without, in this case, needing to multiply or divide by any small musical intervals or commas.  This further leads to the simpler and more easily memorized approximation for the Gibbons-Hawking entropy of the asymptotic empty de Sitter spacetime toward which our part of the universe appears to be headed, $S_\Lambda = 3\pi/\Lambda \approx 5^3 2^{400} \approx 3.23\times 10^{122}$.  One can then work backwards from this to get $\Lambda \approx 3\pi\,5^{-3}2^{-400}$ in Planck units, with the current uncertainty in the cosmological constant (about 3.2\%) sufficient in this case to avoid needing to multiply by any small comma to get agreement with observations.

It will be interesting to see how many years pass before $e$, $m_p$, $m_e$, $t_0$, and $\Lambda$ are known with sufficiently improved accuracy that the mnemonic approximations above, which are just that and are not to be interpreted to have any fundamental significance for their present agreement with observations, will need to be replaced by improved approximations in order to fit the new data.

We can further write various observed Solar System parameters in both conventional units and in Planck units:

Solar mass $M_\odot = 1.988\,49(9)\times 10^{30}$ kg $= 1476.625\,12$ m $(c^2/G)$\\ $ = 9.136\,24(21)\times 10^{37} = 0.536\,980(12) \times 2^{127} = 0.539\,582(12)\, m_p^{-2}$.

It is interesting that $e^{-114}\Lambda = 1.837(59)$ is within its uncertainty the reciprocal of the quantity $M_\odot m_p^2 = 0.539\,582(12)$, so that one can write $\Lambda M_\odot m_p^{2} \approx e^{114}$, as well as $\Lambda M_\odot \approx 2^{127} e^{114}$.

Solar radius $R_\odot = 6.957\times 10^8$ m\\ $= 4.304\times 10^{43}$.

Solar surface temperature $T_\odot = 5772$ K\\ $= 4.074\times 10^{-29}$.

Solar photon luminosity $L_\odot = 3.828\times 10^{26}$ W\\ $ = 1.055\times 10^{-26}$.

Earth mass $M_\oplus = 5.9724(3)\times 10^{24}$ kg 
$= 0.004\,436\,028\,290(9)$ m $(c^2/G)\\ = 2.744\,06(6)\times 10^{32}$.

Earth radius $R_\oplus = 6.371\times 10^6$ m\\ $= 3.942\times 10^{41}$.

Earth density $\rho_\oplus = 5514$ kg/m$^3$\\$ = 1.0696\times 10^{-93}$.

Earth standard gravity $g_\oplus \equiv 9.806\,65$ m/s$^2 = 1.032\,295\, c$/(one year)\\ $= 1.763\,53(4)\times 10^{-51}$.

Astronomical unit 1 au $\equiv 149\,597\,870\,700$ m\\
$ = 499.004\,783\,838$ light seconds\\ $= 9.255\,98(21)\times 10^{45}$.

One day $t_d = 24\times 60\times 60$ s $= 86\,400$ s\\ $= 1.602\,624(37)\times 10^{48}$.

One `month' (orbital period of the Moon) $t_m = 27.321\,661$ days $= 2\,360\,591.5$ s\\ $= 4.378\,63(10)\times 10^{49}$.

One (Julian) year = 365.25 days $= 31\,557\,600$ s\\ $ = 5.853\,58(13)\times 10^{50}$.

Earth precession period $= 25\,771.575\,34$ yr\\ $= 1.508\,560(35)\times 10^{55}$.

\section{Approximate Anthropic Formulas for the Mass and Charge of the Proton and Electron}

Brandon Carter \cite{Carter1970,Carter1974,Carter2007}, who first used the phrase ``anthropic principle'' around 1973, has argued that our existence as observers is favored by certain ranges of the parameters of physics and astronomy, so that if there is a ensemble of many different sets of values, we might expect to observe the parameters to be within favored ranges.  He calculated many such favored values that I shall use here, along with the results of others such as Bernard Carr and Martin Rees \cite{Carr-Rees}, William Press \cite{Press}, and Press and Alan Lightman \cite{PressLightman}.  These predicted that the proton mass $m_p$ and the electron mass $m_e$ should be approximately equal to definite powers of the elementary charge $e$ (the charge of the proton, which is assumed to be the negative of the charge of the electron, as predicted by certain Grand Unified Theories).

In order that nuclei apparently necessary for life as we know it to exist, Carr and Rees \cite{Carr-Rees}, following similar suggestions by Carter \cite{Carter1970,Carter1974,Carter2007}, showed that one needs $m_e/m_p \sim 10\, e^4$.  To avoid numerical factors like 10, I shall replace it by its approximate equivalent $e^{-1}$, giving $m_e \sim m_p e^3$.

Carter \cite{Carter1970,Carter1974,Carter2007}, and also Carr and Rees \cite{Carr-Rees}, argued that life might require (or at least be favored by) the existence both of stars that transmit their energy outward by convection (which might be favorable for planetary formation) and by stars that transmit their energy outward by radiative transfer (which is favorable for the formation of supernovae that produce the heavier elements apparently needed for life).  This leads to the condition that $m_p^3 \sim m_e^2 e^{12}$.

Combining these two conditions leads to $m_p \sim (m_e/m_p)^2 e^{12} \sim (e^3)^2 e^{12} = e^{18}$ and then $m_e \sim e^{18} e^3 = e^{21}$.  Note that these anthropically predicted exponents are very close to the empirical exponents given earlier.

A renormalization group analysis \cite{Marciano:1981un} with $n_g = 3$ generations of quarks and leptons and $N_H = 2$ relatively light Higgs doublets in low energy $SU(3)\times SU(2)\times U(1)$ shows that the inverse coupling constants run approximately linearly with the logarithm of the energy, with calculated coefficients, between the weak scale and the unification scale.  By making the approximations that the proton mass (set by the point where the $SU(3)$ coupling becomes large) is logarithmically near the weak scale (the mass of the W boson), and the unification scale is logarithmically near the Planck scale, I was able to derive the relation \cite{DNP}

$\alpha^{-1} = e^{-2} \approx -(10/\pi)\ln{m_p}$.

Then setting $m_p \sim e^{18}$ leads to an equation approximately determining $e$:

$e^2\ln{e} \approx -\pi/180$.

An interesting mnemonic is that the right hand side is the negative of the number of radians in a degree.  (However, surely this is just a coincidence, depending on the historical accident that the Babylonians divided a circle into 360 degrees, using 60 as a humanly convenient counting number {\it perhaps} arising from the product of the number of fingers on one human hand and the number of phalanxes on the non-thumb fingers of the other hand).

Now let us define anthropic approximations (with subscript $a$) as quantities exactly obeying these approximate equations.  For example, define the anthropic estimate for the elementary charge as $e_a$, the smaller root of the equation

$e_a^2\ln{e_a} = -\pi/180$.

Then one gets

$e_a = 0.083\,927\,766\,8145 = 0.98248\, e$.

Therefore, the anthropic estimate for the elementary charge, which did not use as input {\it any} parameter with a potentially continuous range, agrees within 1.8\% with the actual observed elementary charge.

Although it is not hard to solve the equation for $e_a$ using Newton's method even just on a pocket calculator that can calculate logarithms, for mental calculations the following approximations might be memorable and useful:

$e_a \approx \frac{7}{80} - \frac{1}{280} = \frac{47}{560} \approx 10^{-1.0761} \approx 10^{-\frac{99}{92}}$.

One can then use the Carter-Carr-Rees relations  \cite{Carter1970,Carter1974,Carter2007,Carr-Rees} $m_p \sim e^{18}$ and $m_e \sim e^{21}$ to get anthropic estimates for the masses of the proton and electron:

$m_{pa} \equiv e_a^{18} = 4.2688\times 10^{-20} = 0.55547\, m_p$,

$m_{ea} \equiv e_a^{21} = 2.5236\times 10^{-23} = 0.60295\, m_e$.

These have relative errors of the order of a factor of 2, which is not surprising in view of the large exponents that result in these quantities being many orders of magnitude smaller than the Planck units.  For such quantities far from unity in Planck units, it is probably more meaningful to look as the relative error of the logarithms of the various quantities:

$\ln{e_a}/\ln{e} = 1.007185$, so the logarithm of this anthropic estimate for the elementary charge has a relative error of only 0.7\%.

$\ln{m_{pa}}/\ln{m_p} = 1.013359$, with a relative error of only 1.3\%, and

$\ln{m_{ea}}/\ln{m_e} = 1.009818$, with a relative error of only 1.0\%.

Therefore, by combining the Carter-Carr-Rees power law relations between the elementary charge and the masses of the proton and electron with my renormalization group logarithmic relation, one can get the logarithms of these three quantities to agree with observation with an average relative error of only 1\%.  Note that the simple equations giving these quantities do not depend on any measured parameter that might have a continuous range, though they do depend on discrete integer parameters, such as the dimension of space for the Carter-Carr-Rees relations and the number of generations and of Higgs doublets for my renormalization group relation.

\section{Approximate Anthropic Estimates for Other Physical and Astronomical Parameters}

One can, from the anthropic estimates $e_a$, $m_{pa}$, and $m_{ea}$ for $e$, $m_p$, and $m_e$, readily get estimates of other atomic constants, such as for the Bohr radius $a_0 \equiv e^{-2} m_e^{-1} = 3.2741\times 10^{24}$ in Planck units:

$a_{0a} \equiv e_a^{-2} m_{ea}^{-1} \equiv e_a^{-23} = 5.6256\times 10^{24} = 1.71819\, a_0$,

$\ln{a_{0a}}/\ln{a_0} = 1.009589$.

This has a relative error of only 1.0\% for the logarithm.  One can further get an estimate for the density of liquid water as

$\rho_{H_2Oa} = m_{pa}a_{0a}^{-3} = e_a^{87} = 2.3977\times 10^{-94} = 1.23609\, \rho_{H_2O}$,

$\ln{\rho_{H_2O}}/\ln{\rho_{H_2Oa}} = 1.000983$, with the very small relative error 0.10\%.

Carter and others have shown that stellar masses should be within a couple of orders of magnitude on either side of the Landau mass $M_L = m_p^{-2}$ and hence contain a number of nucleons close to the Landau number $N_L = m_p^{-3}$.  Inserting the anthropic estimate for $m_p$ gives

$M_{*a} \equiv m_{pa}^{-2} \equiv e_a^{-36} = 5.4878\times 10^{38} = 6.0066\, M_\odot$, with

$\ln{M_{*a}}/\ln{M_\odot} = 1.020511$.

For such a star, estimates of the radius, temperature, and luminosity are

$R_{*a} \equiv e_a^{-5} m_{pa}^{-2} \equiv e_a^{-41} = 1.3179\times 10^{44} = 3.0616\, R_\odot$, 

$T_{*a} \equiv e_a^{26.5} = 3.0444\times 10^{-29} = 4313$ K $= 0.7473\, T_\odot$,

$L_{*a} \equiv e_a^{24} = 1.4919\times 10^{-26} = 5.4131\times 10^{26}$ W $= 1.4141\, L_\odot$. 

This gives

$\ln{R_{*a}}/\ln{R_\odot} = 1.013710$,

$\ln{T_{*a}}/\ln{T_\odot} = 1.004456$,

$\ln{L_\odot}/\ln{L_{*a}} = 1.005826$.

The fraction of a star's mass-energy that is converted to radiation during its lifetime is roughly $m_e/m_p \sim e^3$, which is very roughly the nuclear binding energy per nucleon of the heavier elements produced by stellar nuclear burning.  Therefore, the total energy converted to radiation by a star of mass $M_*$ is roughly $M_* m_e/m_p \sim m_p^{-2} m_e/m_p = m_e m_p^{-3} \sim e^{-33}$.

Then a typical star with luminosity $L_* \sim e^{24} \sim m_e^2 m_p^{-1}$ has a lifetime $t_* \sim M_*(m_e/m_p)/L_* \sim m_e^{-1} m_p^{-2} \sim e^{-57}$.  This is precisely the same power of $e$ that we found was an excellent approximation for the age of the universe when multiplied by the magic comma that is $3125/3072 \approx 1.0172526$.

One might expect that typical observers would exist at rather random times during the lifetime of the star that supports them, so that a typical observed age of the universe would be $t_0 \sim t_*$.  Therefore, let us take an anthropic estimate of the observed age of the universe to be

$t_{0a} \equiv m_{ea}^{-1} m_{pa}^{-2} \equiv e_a^{-57} = 2.1746\times 10^{61} 
= 37.15$ Gyr $= 2.6920\, t_0$.

The factor of 2.6920 is mainly due to the factor $(e/e_a)^{57} = (1.01783410136)^{57} = 2.738979912$; even though the anthropic estimate $e_a$ for the elementary charge is within 1.8\% of the observed elementary charge, raising it to a power of large magnitude, such as $-57$, does lead to a relative error of the anthropic estimate for the age of the universe that is a bit more than a factor of 2.  However, on a logarithmic scale, the disagreement is much less:

$\ln{t_{0a}}/\ln{t_0} = 1.007061$, with relative difference only 0.7\%.

Note that this anthropic argument gives a partial explanation for the exponent of the elementary charge $e$ that occurred in the empirical mnemonic approximation for the age of the universe as the magic comma divided by the 57th power of the elementary charge, but it does not explain why one only needs to multiply this power by a factor as close to unity as the magic comma.  The latter fact is surely just a numerical coincidence.

One might note that Raphael Bousso, Lawrence Hall, and Yasunori Nomura \cite{BHN} have predicted that both $t_\Lambda$ and $t_{\mathrm obs}$ (as well as the times of galaxy structure formation and galaxy cooling) should be roughly $\alpha^2/(m_e^2 m_p)$, which with my anthropic estimates comes out to be $e_a^{-56}$, which is one power of $e_a$ smaller than my estimates above, but very close on a logarithmic scale.

Another cosmological parameter, presumably a constant of physics in our part of the universe, is the energy density of dark energy, which has an approximate observed value of

$\rho_\Lambda = \Lambda/(8\pi) \approx (3/8)5^{-3}2^{-400} = 1.16178\times 10^{-123} \approx 0.00538$ erg/m$^3 \approx 3.36$ GeV/m$^3 \approx 3.58\, m_p/$m$^3$, or about the rest-mass energy of 3.6 protons per cubit meter.

(The total mass density of the present universe is about 13/9 times this, or about 5.2 protons per cubic meter \cite{cosmicmnemonics}.)

One might further expect the energy density of the dark energy to be $\rho_\Lambda \sim 1/t_0^2$, so an anthropic value is

$\rho_{\Lambda a} \equiv t_{0a}^{-2} = e_a^{114} = 2.1147\times 10^{-123} = 1.8202\, \rho_\Lambda$,

$\ln{\rho_\Lambda}/\ln{\rho_{\Lambda a}} = 1.002120$.

The logarithm of this apparent constant of physics thus has a relative error of only about 0.2\%.  (More strictly, I should say the real part of the logarithm, since anthropic arguments so far do not predict the sign of the dark energy density, so it would be anthropically acceptable for the logarithm to include an additive term of $\pi i$.)

A crude anthropic estimate for the 4-volume of the observable universe, the part within our past light cone, is

4-volume $\sim e_a^{-228} = 2.236\times 10^{245}$.

Furthermore, a crude anthropic estimate for the number of stars within the observable universe is

Number of stars $\sim N_{*a} \equiv t_{0a}/M_{*a} \equiv m_{ea}^{-1} \equiv e_a^{-21} = 3.963\times 10^{22} = 0.132(3\times 10^{23})$.

The last factor in parentheses, $3\times 10^{23}$, is an observational estimate for the number of stars in the observable universe.  One reason that it is over 7 times larger than the anthropic estimate is that there are more stars of smaller mass, so that the average mass of a star is rather less than the Landau mass $m_p^{-2}$.

One can also give a crude anthropic estimate for the gravitational wave strain from the collision of two black holes of the order of the Landau mass $\sim M_{*a} \equiv m_{pa}^{-2} \equiv e_a^{-36}$ at cosmological distances $\sim t_{0a} \equiv m_{ea}^{-1} m_{pa}^{-2} \equiv e_a^{-57}$, since the strain is of the order of unity near the black holes, at a distance of the order of $M_{*a}$, but then decreases inversely with the first power of the distance $r$ as the gravitational waves propagate away from the source:

Gravitational wave strain $\sim h_a \equiv M_{*a}/t_{0a} \equiv m_{ea} \equiv e_a^{21} = 2.534\times 10^{-23}$.

It is interesting that the the crude anthropic estimate for the gravitational wave strain from colliding black holes across the universe agrees with the crude anthropic estimate for the reciprocal of the number of stars in the observable universe, and that this estimate is the same as that for the mass of the electron in Planck units, which is approximately the 21st power of the elementary charge.

\section{Approximate Anthropic Estimates for Properties of\\Biology and Habitable Planets}

Now let us focus on biological properties and properties of habitable planets.

Observed normal human body temperature is about 37 C or

$T_h = 310.15$ K $= 2.1891\times 10^{-30}$.

Press \cite{Press}, and also Press and Lightman \cite{PressLightman}, have estimated biological habitable temperatures (not so low that they freeze and not so high that they cook) as 

$T_h \sim 0.1(m_e/m_p)^{1/2} m_e e^4 \sim e^{27.5}$.

Let us therefore take an anthropic estimate for the temperature of the Earth as a habitable planet as

$T_{\oplus a} \equiv e_a^{27.5} = 2.555\times 10^{-30} = 366.0$ K $= 1.167 T_h$.

This is about 88 C, which is bit too hot for humans (though not for some life on Earth), but in Planck units the logarithm is only off from that of the observed normal human body temperature by about 0.2\%:

$\ln{T_h}/\ln{T_{\oplus a}} = 1.002269$.

Press \cite{Press}, and Press and Lightman \cite{PressLightman}, used the requirement that we breathe an evolved planetary atmosphere, meaning that at the temperature $T_h$, hydrogen will have mostly escaped from the planet, but not heavier gases such as oxygen.  This leads to a gravitational potential at the surface of the planet of mass $M$ and radius $R$ to be $M/R$ that is just a few times $T_h/m_p \sim e^{9.5}$, say $M/R \sim e^9$.  Then one notes that $M/R^3 \sim \rho_{H_2O} \sim m_{p}a_{0}^{-3} \sim e^{87}$.  Solving these relations gives anthropic estimates for the radius, mass, acceleration of gravity, and satellite orbital speed of a habitable planet as

$R_{\oplus a} \equiv m_{pa}^{-1}m_{ea}^{-1} \equiv e_a^{-39} = 9.283\times 10^{41} = 15\,003$ km $= 2.355 R_\oplus$,

$\ln{R_{\oplus a}}/\ln{R_\oplus} = 1.008943$,

$M_{\oplus a} \equiv m_{pa}^{-4}m_{ea}^{2} \equiv e_a^{-30} = 1.918\times 10^{32} = 0.6989 M_\oplus$,

$\ln{M_\oplus}/\ln{M_{\oplus a}} = 1.004819$,

$g_{\oplus a} \equiv M_{\oplus a}/R_{\oplus a}^2 \equiv m_{pa}^{-2}m_{ea}^{4} \equiv e_a^{48} = 2.226\times 10^{-52} = 0.1262 g_\oplus$,

$v_{\oplus a} \equiv \sqrt{M_{\oplus a}/R_{\oplus a}} \equiv (m_{ea}/m_{pa})^{3/2} \equiv e_a^{4.5} = 0.5447 v_\oplus$.

Press and Lightman \cite{PressLightman} estimated the length of the day, $t_d = 86\,400$ s $= 1.603\times 10^{48}$, as coming from the maximum rotation rate a planet can have without breaking up, since planets generally form with angular velocity near this maximum, which is $\omega \sim (G\rho)^{1/2} \sim (G m_p/a_0^3)^{1/2} = (G m_p m_e^3 e^6)^{1/2} \sim e^{43.5}$.  The length of the day is then $t_d = 2\pi/\omega \sim e^{-44}$ if we round to the nearest whole power of the elementary charge $e$, so let us take the anthropic estimate for the length of the day to be

$t_{da} \equiv e_a^{-44} = 2.229\times 10^{47} = 200.3$ minutes $= 0.1391$ day,

$\ln{t_d}/\ln{t_{da}} = 1.018093$.

This estimate for the length of the day is of course too low because the Earth has lost a significant fraction of its spin angular momentum, most particularly by tidal friction that transfers angular momentum to the Moon, and perhaps early in the Earth-Moon evolution, also to the orbital angular momentum of the Earth-Moon system around the Sun.

For the Earth surface to be at temperature $\sim T_h \sim T_{\oplus a} \equiv e_a^{27.5}$ while the surface of the Sun is at temperature $\sim T_{*a} \equiv e_a^{26.5}$, a temperature greater by a factor of $e_a^{-1} = 11.915$ (within a factor of 2 of the actual temperature ratio $5772/288 = 20.04$), the Earth should be at a distance $r \approx (1/2)R_\odot(T_\odot/T_\oplus)^2$.  Dropping the factor of 1/2 and using the anthropic estimates $R_\odot \sim R_{*a} \equiv e_a^{-41}$ and $T_\odot/T_\oplus \sim T_{*a}/T_{\oplus a} \equiv e_a^{-1}$ then gives an anthropic estimate of the conventional value of the semimajor axis of the Earth's orbit, the astronomical unit, 1 au $\equiv 149\,597\,870\,700$ m\\
$= 9.255\,98(21)\times 10^{45}$, as

$r_{\oplus a} \equiv e_a^{-43} = 1.8709\times 10^{46} = 2.0213$ au,

$\ln{r_{\oplus a}}/\ln{(\mathrm{au})} = 1.006649$.

This is coincidentally a case in which if I had left in the factor of 1/2, the anthropic estimate would have been 1.0107 au, with only a 1\% error for the actual value, but I am avoiding keeping factors of 1/2 and am only trying to get the right power of the anthropic estimate $e_a = 0.083\,927\,766\,8145$ of the elementary charge.

From the anthropic estimate of the stellar mass as $M_{*a} \equiv m_{pa}^{-2} \equiv e_a^{-36}$ and of the Sun-Earth distance (the astronomical unit) as $r_{\oplus a} \equiv e_a^{-43}$, we get an anthropic estimate for the length of the year as $2\pi/\omega_o$ with orbital angular velocity obeying $\omega_o^2 \sim M_\odot/(1\ \mathrm{au})^3 \sim M_{*a}/r_{\oplus a}^3 \equiv e_a^{93} \equiv m_{pa}^4 m_{ea}$.  Then approximating $2\pi$ by $e_a^{-1/2}$ gives the anthropic estimate for the length of the year (e.g., the Julian year = yr = 365.25 days $= 31\,557\,600$ s $= 5.8536\times 10^{50}$ that is conventionally used in astronomy) as

$t_{\oplus a} \equiv e_a^{-47} = 3.7708\times 10^{50} = 235.29$ days $= 0.64419$ yr,

$\ln{\mathrm{yr}}/\ln{t_{\oplus a}} = 1.003776$.

It is not at all obvious that a planet needs a moon to have life and observers, though sometimes it is suggested that tides may help the evolution of sea life near ocean shores.  Also, I have no prediction for the Moon-Earth mass ratio $M_m/M_\oplus$ that we observe to have the memorable value .0123.  But one can work out a crude prediction for the length of the month (taken here to be the orbital period of the Moon around the Earth, $t_m = 27.321\,661$ days $= 2\,360\,591.5$ s $=4.3786\times 10^{49}$ Planck times) from the age of the solar system, $t_{SS} = 4.5682(3)$ Gyr $= 2.6740\times 10^{60}$, the mass $M_\oplus$ and radius $R_\oplus$ of the Earth, and the Moon-Earth mass ratio $M_m/M_\oplus$.  Taking this mass ratio to be significantly smaller than unity, as it is for our Earth-Moon system, $t_m \approx 2\pi \sqrt{r^3/M_\oplus}$ in terms of the Earth-Moon separation distance $r$.

The idea is that as the Moon orbits the Earth, it provides tidal friction and torque on the Earth that increases the Moon's orbital angular momentum $L = M_m r v = M_m \sqrt{M_\oplus r}$, where $v = \sqrt{M_\oplus/r}$ is the Moon's orbital velocity under the simplifying assumption that the orbit is circular.

The tidal force per mass of the Moon on masses on opposite sides of the Earth, separated by $2R_\oplus$, is $4R_\oplus GM_m/r^3$.  Since I am using Planck units, I shall continue to set $G=1$.  This tidal force raises a tide on the Earth by a height of the order of the radius of the Earth multiplied by the ratio of this tidal force to the Earth's acceleration of gravity, $g \approx GM_\oplus/R_\oplus^3$, which is $h \sim (M_m/M_\oplus)(R_\oplus/r)^3 R_\oplus$.  The amount of mass in this tide will be of the order of $(M_m/M_\oplus)(R_\oplus/r)^3 M_\oplus = M_m (R_\oplus/r)^3$.

Assuming that the spin angular velocity of the Earth is sufficiently greater than the orbital angular velocity $\sqrt{M_m/r^3}$ of the Moon that the line between high tides on the opposite sides of the Earth is twisted by an angle of the order of unity (in radians) from the direction to the Moon, then the torque exerted by the Moon will be of the order of the tidal force of the Moon on the tide, multiplied by the radius of the Earth, or $(R_\oplus M_m/r^3)[M_m (R_\oplus/r)^3]R_\oplus$.  

The time it has taken for the Moon to move out to its present radius, which is very nearly $t_{SS}$ since the Moon apparently formed shortly after the solar system did, is roughly the orbital angular momentum of the Moon divided by this torque, or 

$t_{SS} \sim \frac{M_m \sqrt{M_\oplus r}}{(R_\oplus M_m/r^3)[M_m (R_\oplus/r)^3] R_\oplus} = \frac{M_\oplus}{M_m}\left(\frac{r}{R_\oplus}\right)^5\sqrt{\frac{r^3}{M_\oplus}} \sim \frac{M_\oplus}{M_m}\left(\frac{r}{R_\oplus}\right)^5 t_m \approx \frac{1}{0.0123}(60.3357)^5$ (0.074\,802\,631 yr) = 4.8628 Gyr $= 1.0645 t_{SS}$, fortuitously within 7\% of the correct answer.

If one writes $t_{SS} \sim \frac{\rho_\oplus}{\rho_m}\left(\frac{R_m}{R_\oplus}\right)^2 \left(\frac{r}{R_m}\right)^5 t_m$, all but the first factor can be determined by observations of the Moon's angular size in the sky, the curvature of the Earth's shadow on the Moon during a lunar eclipse, and the length of the month.  Then if one just assumes that the ratio of densities is of the order of unity, one gets an estimate for the age of the solar system that is 2.95 Gyr, which is about 65\% of the actual age but again within a factor of 2 of the correct answer.

If in $t_{SS} \sim \frac{M_\oplus}{M_m}\left(\frac{r}{R_\oplus}\right)^5 t_m$ we replace $r$ by $(M_\oplus t_m^2)^{1/3}$ and $M_\oplus/r_\oplus^3$ by $\rho_\oplus$, we get $t_{SS} \sim (M_\oplus/M_m)\rho_\oplus^{5/3}t_m^{13/3}$, which one can solve for the length of the month as $t_m \sim (M_m/M_\oplus)^{3/13}\rho_\oplus^{-5/13}t_{SS}^{3/13}$.  Then assuming that $(M_m/M_\oplus)^{3/13}$ is of the order of unity (it is 0.36242 for the Earth-Moon system) and using the anthropic estimates $\rho_\oplus \sim \rho_{H_2Oa} \equiv m_{pa}a_{0a}^{-3} \equiv m_{pa}^{-1}m_{ea}^5 \equiv e_a^{87}$ and $t_{SS} \sim t_{0a} \equiv m_{ea}^{-1} m_{pa}^{-2} \equiv e_a^{-57}$ gives an anthropic estimate for the length of the month as 

$t_{ma} \equiv \rho_{H_2Oa}^{-5/13}t_{0a}^{3/13} \equiv m_{pa}^{-1/13}m_{ea}^{-28/13} \equiv e_a^{-606/13} = e_a^{-46.6154} = 1.4540\times 10^{50} = 90.7235$ days $= 3.3206\, t_m$,

$\ln{t_{ma}}/\ln{t_m} = 1.010500$.

If we had inserted the factor $(M_m/M_\oplus)^{3/13} = 0.36242$ for the Earth-Moon system, which so far as I know is not determined by anthropic reasoning, then one gets  32.8798 days $= 1.2034 t_m$.

The spin of the Earth precesses because of the tidal torque by the Moon and the Sun, with a precession time $t_p = 25\,771.575\,34$ years that may be estimated by dividing the spin angular momentum of the Earth, $L_\oplus \sim M_\oplus R_\oplus^2 \omega_\oplus$, by the torque from the Moon (the main source), which is $\tau \sim f M_\oplus M_m R_\oplus^2/r^3$ with Earth-Moon distance $r$ as before, and with $f = 1/298.257\,223\,563$ being the flattening factor for the Earth whose order of magnitude is $f \sim \omega_\oplus^2/\rho_\oplus$.  Using the length of the day as $t_d \approx 2\pi/\omega_\oplus$ and dropping factors like $2\pi$ then gives

$t_p \sim \frac{\rho_\oplus}{\rho_m}\left(\frac{r}{R_m}\right)^3 t_d = 48\,956$ yr $= 1.900 t_p$.

Adding the tidal effect of the Sun would reduce the precession time by a factor of 0.68514 to 33\,542 yr $= 1.3015 t_p$.  If one ignored the effect of the Sun and also the ratio of the densities of the Earth and Moon, and wrote $R_\oplus/r = \theta_m \approx 1/220$ as the average angular radius of the Moon as seen from Earth, then one gets

$t_p \sim t_d/\theta_m^3 \approx 220^3$ days $= 10\,648\,000$ days $= 29\,153$ years $= 1.13\, t_p$.

Thus just from simple observations (but not anthropic reasoning), one can estimate the Earth's precession period as the length of the day divided by the cube of the angular radius of the Moon in the sky.

To estimate the Earth's precession period from anthropic considerations rather than from observations, return to

$t_p \sim \frac{\rho_\oplus}{\rho_m}\left(\frac{r}{R_m}\right)^3 t_d = \frac{M_\oplus}{M_m}\left(\frac{r}{R_\oplus}\right)^3 t_d \sim \frac{M_\oplus}{M_m}\rho_\oplus t_m^2 t_d$.

Now if we drop the $M_\oplus/M_m$ factor that is about 81.3 for the Earth-Moon system but which I do not know how to estimate anthropically, and if we use our previous estimates $\rho_\oplus \sim \rho_{H_2Oa} \equiv e_a^{87}$, $t_m \sim t_{ma} \equiv e_a^{-606/13}$, and $t_d \sim t_{da} \equiv e_a^{-44}$, then we get the following anthropic estimate for the Earth's precession period:

$t_{pa} \equiv \rho_{H_2Oa} t_{ma}^2 t_{da} \equiv e_a^{-653/13} = e_a^{-50.2308} = 1.1299\times 10^{54} = 1930.3$ yr $= 0.074900\, t_p$,

$\ln{t_p}/\ln{t_{pa}} = 1.020822$.

The relative error for $t_{pa}$ is larger than for $t_{ma}$ because $M_\oplus/M_m$ comes in with exponent $+1$ for $t_p$ rather than with the exponent $-3/13$ that is much closer to zero that it has for $t_m$.  In fact, including the factor $M_\oplus/M_m \approx 81.3$ for the Earth-Moon system would make $(M_\oplus/M_m) t_{pa} = 6.089\, t_p$, now too large.

\section{Approximate Anthropic Estimates for the Tallest\\Running Animal}

Press and Lightman \cite{PressLightman} estimate the peak power output of an animal of temperature $T$ and height $h$ to be limited by the cooling rate which they estimate as $P \sim CTh$, where they estimate the conductivity as $C \sim m_e^2 e^6 (m_e/m_p)^{1/2}$.  The anthropic estimates of Carter \cite{Carter1970,Carter1974,Carter2007} and of Carr and Rees \cite{Carr-Rees} would then make $C \sim e^{49.5}$.  Since this is just an order-of-magnitude estimate, for later convenience when combined with other anthropic estimates, I shall round the exponent down to 49 and define my anthropic estimate for the conductivity to be $C_a \equiv e_a^{49}$.

Now I shall equate this to the power expended during running for the tallest land animal (e.g., a giraffe) \cite{giraffe}, which for an animal of mass $m \sim \rho_{H_2O} h^3$ and height $h$ running at speed $v$ Press and Lightman estimate to be $P \sim mv^3/h$, using up energy $\sim mv^2$ during each stride time $\sim h/v$.  Going beyond Press and Lightman \cite{PressLightman}, I assume that the running speed $v$ is roughly the speed of a pendulum of length $h$ undergoing large-amplitude oscillations, so $v \sim \sqrt{g_\oplus h}$. 

Today the tallest running animal on Earth is a giraffe, for which the tallest recorded had a height $h_g = 5.88$ m $=3.64\times 10^{35}$.

Equating $P \sim m g_\oplus^{3/2} h^{1/2} \sim \rho_{H_2O} g_\oplus^{3/2} h^{7/2}$ to $P \sim CTh$ then gives $h^{5/2} \sim T C \rho_{H_2O}^{-1} g_\oplus^{-3/2} \sim e_a^{27.5+49-87-72} = e_a^{-82.5}$, which leads to the following anthropic estimate for the height of the tallest running land animal:

$h_a \equiv e_a^{-33} = 3.2442\times 10^{35} = 5.2434$ m $= 0.892 h_g$,

$\ln{h_g}/\ln{h_a} = 1.001401$,\\
so the logarithm is correct to within one part in 700.

Using $h_a \equiv e_a^{-33}$ and $g_{\oplus a} \equiv e_a^{48}$ gives an estimate of the stride time and running speed for the tallest running animal as

$t_a \equiv \sqrt{h_a/g_{\oplus a}} \equiv e_a^{-40.5} = 3.8179\times 10^{43} = 2.058$ s,

$v_a \equiv \sqrt{g_{\oplus a} h_a} \equiv e_a^{7.5} = 8.4975\times 10^{-9} = 2.5475$ m/s = 9.1710 km/hr.

This is perhaps about the speed an old theorist like me can run for a short period of time, though even if I could maintain this speed for a long time, it would take 4.6009 hours ($3.0723\times 10^{47}$ Planck times) to run a marathon, whose length is exactly 26 miles and 385 yards, or 138\,435 feet, or 42.194\,988 kilometers, or $2.610\,706(61)\times 10^{39}$ Planck units.  This time of 4.6009 hours is 2.2453 times the best time of 2 hours 2 minutes 57 seconds, or 7377 seconds, or 2.049\,167 hours, or $1.368\,351(32)\times 10^{47}$ Planck times.

One can also get an anthropic estimate for the specific power or power per mass of the tallest running animal as

$p_a \equiv v_a^3/h_a \equiv g_{\oplus a}^{3/2} h_a^{1/2} \equiv e_a^{55.5} = 1.8913\times 10^{-60} = 3.1530$ W/kg = 65.110 kilocalories/day/kg,
which is about 6.5 times a measurement \cite{REE} of 9.99 kilocalories per day per kilogram of the derivative of the resting energy expenditure in healthy
human individuals with respect to mass (at fixed height, age, and sex).  

It is interesting that this anthropic estimate of the specific power of the tallest land animal while running is just a factor of $e_a^{-1.5} = 41.128$ times the anthropic estimate of the Hubble expansion rate of the universe, $t_{0a}^{-1} \equiv e_a^{-57} = 1.5831$ kilocalories/day/kilogram, which is about 16\% of the 9.99 kilocalories/day/kilogram mentioned above.  If one takes instead the measured value of the Hubble constant, $H_0 = 67.8(9)$ km/s/Mpc, this corresponds to 0.198(3) W/kg.

Thus a convenient unit for the specific power of humans might be the {\it{hubble}}, which could be defined as the square of the speed of light multiplied by some nominal value of the Hubble constant, such as one that gives

1 hubble $\equiv 0.2$ W/kg $\approx 4.130\,019\,120\,46$ kilocalories/day/kilogram.

Then the hubble would be a unit of specific power is within the range of typical values for humans and other large animals.  For example, if it went with 100\% efficiency into climbing (gaining altitude at constant velocity) with the Earth standard acceleration of gravity $g_\oplus \equiv 9.806\,65$ m/s$^2$, then one hubble would correspond to a climbing rate of 73.4196 meters per hour, or 240.878 feet per hour.

\section{Conclusions}

Therefore, if we take the two anthropic power-law relations Brandon Carter \cite{Carter1970,Carter1974,Carter2007}, and Bernard Carr and Martin Rees \cite{Carr-Rees}, found between the elementary charge $e$ and the masses of the proton $m_p$ and electron $m_e$, and combine them with the logarithmic renormalization group relation I found \cite{DNP}, one gets unique anthropic estimates for the elementary charge $e$ as the smaller solution of $e_a^2\ln{e_a} = -\pi/180$, namely $e_a = 0.083\,927\,766\,8145 \approx 47/560 \approx 10^{-1.0761} \approx 10^{-99/92}$, for the mass of the proton as $m_{pa} = e_a^{18}$, and for the mass of the electron as $m_{ea} = e_a^{21}$.

Then one can use both old and new physical, astronomical, and anthropic arguments to get estimates of the masses, sizes, and times of such things as atoms, giraffes, the Earth, the Sun, the day, the month, the year, the Earth precession time, the age of the solar system, and the age of the universe.  In Planck units, the estimated logarithms of these quantities are usually within a few percent or less of the observed quantities.

Another proposal \cite{BH,BFLR1,BFRL2} is that the huge size of the universe, and the tiny value of the cosmological constant, is related to the number of vacua in the landscape.  This might be so, but the absolute value of the common logarithm of the cosmological constant is just a bit more than 120, whereas the usual number given as an estimate for the common logarithm of the number of vacua is 500 (which might itself be an underestimate by a large factor).  So at present it appears that the logarithm of the number of vacua may be more than four times the logarithm of the inverse cosmological constant, more than 400 times the error in the estimates of the logarithm of the inverse cosmological constant from the anthropic considerations given in this paper.  Of course, the arguments used in this paper are more complex and depend more specifically upon the observed structure of the effective laws of physics in our part of the landscape than the more generic arguments of the competing proposal, so one might hope that such a simpler explanation would be viable.  However, the success of the present anthropic arguments for giving good estimates for the logarithms of the size of the universe and other structures within it suggests that there may be a strong anthropic weighting toward universes that have the anthropic relations that Carter, Carr, Rees, Press, Lightman, Bousso, Hall, Nomura, and others have discovered, along with the renormalization group properties I have found that allows one to convert those anthropic relations to definite predictions of the size of the observable universe and its parts.

I have benefited from discussions on this subject with Raphael Bousso and Juan Maldacena.  I appreciated the hospitality of the George P.\ and Cynthia W.\ Mitchell Institute for Fundamental Physics and Astronomy of Texas A \& M University. and of the Mitchell family at the Cook's Branch Nature Conservancy, where part of this paper was prepared.  I am also grateful for the hospitality of the Perimeter Institute for Theoretical Physics in Waterloo, Ontario, Canada, where the first part of this paper was written up.  Other parts were prepared during the hospitality of Paul Davies and the Beyond Center of Arizona State University and during the hospitality of Gabor Kunstatter at the Physics Department of the University of Winnipeg.  This research was supported in part by the Natural Sciences and Engineering Research Council of Canada.

%\newpage


\baselineskip 4pt

\begin{thebibliography}{99}

\textwidth 18cm
\textheight 25cm
\topmargin -25mm
\oddsidemargin -7mm
\evensidemargin -7mm

\bibitem{BP} R.~Bousso and J.~Polchinski, ``Quantization of Four Form Fluxes and Dynamical Neutralization of the Cosmological Constant,'' JHEP {\bf 0006}, 006 (2000) [arXiv:hep-th/0004134].

\bibitem{BP2} R.~Bousso and J.~Polchinski, ``The string theory landscape,'' Sci.\ Am.\ {\bf 291}, 60-69  (2004).

\bibitem{KKLT} S.~Kachru, R.~Kallosh, A.~D.~Linde, and S.~P.~Trivedi, ``De Sitter Vacua in String Theory.'' Phys.\ Rev.\ D{\bf 68}, 046005 (2003) [arXiv:hep-th/0301240].

\bibitem{Susskind} L.~Susskind, {\em The cosmic Landscape: String Theory and the Illusion of Intelligent Design} (Little, Brown, New York, 2005).

\bibitem{measure} See references in, for example,  D.~N.~Page, ``Cosmological Measures without Volume Weighting,'' JCAP {\bf 0810}, 025 (2008) [arXiv:0808.0351 [hep-th]].

\bibitem{NIST} ``CODATA Internationally Recommended 2014 Values of the
Fundamental Physical Constants,'' http://physics.nist.gov/cuu/Constants/index.html (retrieved 2017 March 8).

\bibitem{vulture} ``Vulture Comma,'' https://xenharmonic.wikispaces.com/vulture+comma (retrieved 2017 March 8).

\bibitem{landscape} ``Landscape Family,'' http://xenharmonic.wikispaces.com/Landscape+family (retrieved 2017 March 8).

\bibitem{magic} ``Magic Comma,'' http://xenharmonic.wikispaces.com/magic+comma (retrieved 2017 March 8).

\bibitem{Carter1970}  B. Carter, ``Large Numbers in Astrophysics and Cosmology'' (paper presented at Clifford Centennial Meeting, Princeton, 1970).

\bibitem{Carter1974} 
  B.~Carter,
  ``Large Number Coincidences and the Anthropic Principle in Cosmology,'' in M. S. Longair (ed.), {\em Confrontation of Cosmological Theory with Observational Data.   IAU Symposium No.\ 63} (Riedel, Dordrecht, 1974), pp. 291-298; reprinted in Gen.\ Relativ.\ Gravit.\ (2011) {\bf 43}: 3225. doi:10.1007/s10714-011-1258-7

\bibitem{Carter2007} 
  B.~Carter,
  ``The Significance of Numerical Coincidences in Nature,''
  arXiv:0710.3543 [hep-th] (line by line transcript, preserving the original page numbering, of the stencilled preprint issued in 1967).

\bibitem{Carr-Rees}  B. J. Carr and M. J. Rees, ``The Anthropic Principle and the Structure of the Physical World,'' Nature {\bf 278}, 605-612 (1979).

\bibitem{Press} W.~H.~Press,
``Man's Size in Terms of Fundamental Constants,''
Am.\ J.\ Phys.\ {\bf 48}, 597-598 (1980),
http://dx.doi.org/10.1119/1.12326.
  
\bibitem{PressLightman} W.~H.~Press and A.~P.~Lightman,
  ``Dependence of Macrophysical Phenomena on the Values of the Fundamental Constants,''
  Phil.\ Trans.\ Roy.\ Soc.\ Lond.\ A {\bf 310}, 323 (1983),
  doi:10.1098/rsta.1983.0094.
  
\bibitem{Marciano:1981un} 
  W.~J.~Marciano and G.~Senjanovic,
  ``Predictions of Supersymmetric Grand Unified Theories,''
  Phys.\ Rev.\ D {\bf 25}, 3092 (1982),
  doi:10.1103/PhysRevD.25.3092.

\bibitem{DNP} D.~N.~Page, ``Anthropic Estimates of the Charge and Mass of the Proton,'' Phys.\ Lett.\  B{\bf 675}, 398-402  (2009) [arXiv:hep-th/0302051].

\bibitem{BHN} R.~Bousso, L.~J.~Hall, and Y.~Nomura, ``Multiverse Understanding of Cosmological Coincidences,''  Phys.\ Rev.\  D{\bf 80},  063510 (2009)  [arXiv:1001.1155 [hep-th]].

\bibitem{cosmicmnemonics} 
  D.~Scott, A.~Narimani and D.~N.~Page,
  ``Cosmic Mnemonics,''
  Phys.\ Canada {\bf 70}, 258 (2014)
  [arXiv:1309.2381 [astro-ph.CO]].

\bibitem{giraffe} D.~N.~Page, ``The Height of a Giraffe,''  Found.\ Phys.\ {\bf 39}, 1097-1108 (2009) doi:10.1007/s10701-009-9322-9 [arXiv:0708.0573 [hep-th]].

\bibitem{REE} 
M.~D.~Mifflin, S.~T.~St.\ Jeor, L.~A.~Hill, B.~J.~Scott, S.~A.~Daugherty, and Y.~O.~Koh, ``A New Predictive Equation for Resting Energy Expenditure in Healthy Individuals,'' Am.\ J.\ Clin.\ Nutr.\ {\bf 51}, 241-247 (1990).

\bibitem{BH} R.~Bousso and R.~Harnik, ``The Entropic Landscape,''  Phys.\ Rev.\  D{\bf 82}, 123523  (2010) [arXiv:0902.2263 [hep-th]].

\bibitem{BFLR1} R.~Bousso, B.~Freivogel, S.~Leichenauer, and V.~Rosenhaus, ``A Geometric Solution to the Coincidence Problem, and the Size of the Landscape as the Origin of Hierarchy,'' Phys.\ Rev.\ Lett.\ {\bf 106}, 101301 (2011) [arXiv:1011.0714 [hep-th]].

\bibitem{BFRL2} R.~Bousso, B.~Freivogel, S.~Leichenauer, and V.~Rosenhaus, ``Geometric Origin of Coincidences and Hierarchies in the Landscape,'' arXiv:1012.2869 [hep-th].

\end{thebibliography}
\end{document}